\documentstyle[epsfig,natbib2,natbibmnfix]{mn2e}

\newcommand{\be}{\begin{equation}}
\newcommand{\ee}{\end{equation}}

\newcommand{\apj}{ApJ}

\newcommand{\mnras}{MNRAS}
\newcommand{\aap}{A\&A}

\newcommand{\apjl}{ApJL}

\newcommand{\nat}{Nature}
\newcommand{\mbh}{M_{\rm BH}}

\def\ltsima{$\; \buildrel < \over \sim \;$}
\def\simlt{\lower.5ex\hbox{\ltsima}}
\def\gtsima{$\; \buildrel > \over \sim \;$}
\def\simgt{\lower.5ex\hbox{\gtsima}}
\def\sgra{Sgr~A$^*$}

\def\ledd{{L}_{\rm Edd}}

\def\msun{{\,{\rm M}_\odot}}
\def\rsun{{\,R_\odot}}
\def\lsun{{\,L_\odot}}

\def\del#1{{}}

\title[]{Massive stars in sub-parsec rings around galactic centers.}

\author[S.~Nayakshin] {\parbox{18cm}{Sergei Nayakshin}\vspace{0.3cm}\ \\ $^1$
Dept. of Physics \& Astronomy, University of Leicester, Leicester, LE1 7RH,
UK}

\begin{document}

\maketitle

\begin{abstract}
We consider the structure of self-gravitating marginally stable accretion
disks in galactic centers in which a small fraction of the disk mass has been
converted into proto-stars. We find that proto-stars accrete gaseous disk
matter at prodigious rates. Mainly due to the stellar accretion luminosity,
the disk heats up and geometrically thickens, shutting off further disk
fragmentation. The existing proto-stars however continue to gain mass by gas
accretion. As a results, the initial mass function for disk-born stars at
distances $R \sim 0.03-3$ parsec from the super-massive black hole should be
top-heavy. The effect is most pronounced at around $R\sim 0.1$ parsec. We
suggest that this result explains observations of rings of young massive stars
in our Galaxy and in M31, and predict that more of such rings will be
discovered.
\end{abstract}

\renewcommand{\thefootnote}{\fnsymbol{footnote}}
\footnotetext[1]{E-mail: {\tt Sergei.Nayakshin {\em at} astro.le.ac.uk}}

\section{Introduction}
\label{sec:intro}

Accretion disks around super-massive black holes (SMBHs) have been predicted
to be gravitationally unstable at large radii where they become too cool to
resist self-gravity and can collapse to form stars or planets
\citep{Paczynski78,Kolykhalov80,Lin87,Collin99,Gammie01,Goodman03}. There is
now observational evidence that the two rings of young massive stars of size
$\sim 0.1$ parsec in the centre of our Galaxy were formed in situ
\citep{NS05,Paumard05}, confirming the theoretical predictions. In our
neighbouring Andromeda Galaxy (M31), \cite{Bender05} recently discovered a
population of hot blue stars in a disk or ring of similar size, i.e. with
radius of $\sim 0.15$ parsec. The significance of this discovery is that SMBH
in M31 is determined to be as massive as $\mbh \approx 1.4 \times 10^8 \msun$,
or about 40 times more massive than the SMBH in the Milky Way. This fact alone
rules out (Eliot Quataert, private communication) the other plausible
mechanism of forming stellar disks around SMBHs, e.g. the massive cluster
migration scenario \citep[e.g.,][]{Gerhard01}, because the shear presented by
the M31 black hole is much stronger than it is at same distance from \sgra,
and its hard to see how a realistic star cluster would be able to survive that
\citep{Gurkan05}.

In this paper we shall attempt to understand what happens with the gaseous
accretion disk around a SMBH when the disk crosses the boundary of the
marginal stability to self-gravitation \citep{Toomre64} and forms first
stars. We find that in a range of distances from SMBH, interestingly centered
at $R\sim 0.1$ parsec, creation of first low-mass proto-stars should lead to
very rapid accretion on these stars. The respective accretion luminosity
greatly exceeds the disk radiative cooling, thus heating and puffing the disk
up. The new thermal equilibrium reached is that of a disk stable to
self-gravity where further disk {\em fragmentation} is shut off. Star
formation is however continued via accretion onto the existing proto-stars,
which then grow to large masses. We therefore predict that stellar disks
around SMBHs should generically posses top-heavy IMF, as seems to be observed
in \sgra\ \citep{NS05,Nayakshin05b}. In the discussion section we note three
main differences between star formation process in a ``normal'' galactic
environment and that in an accretion disk near a SMBH.

\section{Pre-collapse accretion disk}\label{sec:qeq1}

In this section we determine the structure of the marginally stable accretion
disk, $Q\approx 1$, i.e. the disk structure just before first gravitationally
bound objects form. We envisage a situation in which the disk of a finite
radius has been created by a ``mass deposition event'' on a time scale much
shorter than the disk viscous time, but much longer than the local dynamical
time, $1/\Omega$ (see below). Such an event could be a collision of two large
gas clouds al larger distances from the SMBH, which cancelled most of the
angular momentum of the gas, or cooling of a large quantity of hot gas that
already had a specific angular momentum much smaller than that of the Galaxy
(hot gas can be supported by its pressure in addition to rotation).  In these
conditions, it is reasonable to expect that the disk will settle into a local
thermal equilibrium, in which the gas is heated via turbulence generated by
self-gravitation \citep{Gammie01} and is cooled by radiation. The
magnitude of viscosity $\alpha$-parameter, and the disk
cooling time, $t_{\rm cool}$, are then coupled by
\citep{Gammie01,Levin03b,Rice05}:
\begin{equation}
t_{\rm cool} = \frac{4}{9} \; \frac{1}{\gamma (\gamma-1) \alpha \Omega}
\end{equation}
where $\gamma$ is the adiabatic index of gas. As we shall see below, for the
parameters of interest, the evolution of the disk after star formation is
turned on proceeds on a time scale again shorter than the local viscous
time. Therefore, below we assume that the disk is in the hydrostatic and
thermal equilibrium, but not in a steady accretion state, when the accretion
rate $\dot M(R) =$ const.  We now estimate the conditions in the disk (as a
function of radius $R$) when it reaches surface density large enough to suffer
local gravitational collapse. Star formation is a local process in this
approach, and different rings in the disk could become gravitationally
unstable at different times.

The appropriate accretion disk equations for $Q\sim 1$ have been discussed by
many authors (see references in the Introduction). The hydrostatic balance
condition yields
\begin{equation}
c_s^2 \equiv \frac{P}{\rho} = H^2 \Omega^2\;,
\label{cs}
\end{equation}
where $c_s$ is the isothermal sound speed, $P$ and $\rho$ are the total
pressure and gas density, $H$ is the disk scale height and $\Omega^2 =
(G\mbh/R^3 + \sigma_v^2/R^2)$ is the Keplerian angular frequency at radius $R$
from the black hole.  $\sigma_v$ here is the stellar velocity dispersion just
outside the SMBH radius of influence, i.e. where the total stellar mass
becomes larger than $\mbh$. Using equation \ref{cs}, the disk midplane density
is determined by inversion of the definition of \cite{Toomre64} $Q$-parameter:
\begin{equation}
\rho = \frac{\Omega^2}{\sqrt{2}\pi G Q}\;.
\label{rho}
\end{equation}

To solve for temperature of the disk, we should specify heating and cooling
rates per unit area of the disk. The former is coupled to the rate of the mass
transfer through the disk, $\dot M$:
\begin{equation}
Q^{+}_{\rm d} = \frac{3 \Omega^2 \dot M}{8 \pi}\;.
\label{qdisk}
\end{equation}
The accretion rate is given by
\begin{equation}
\dot M = 3 \pi \nu \Sigma \;,
\label{mdot}
\end{equation}
where $\Sigma = 2 H \rho$ is the disk surface density. The kinematic viscosity
$\nu$ in terms of the \cite{Shakura73} prescription is $\nu = \alpha c_s
H$. Marginally stable self-gravitating disks are believed to have $\alpha\sim
1$ \citep{Lin87,Gammie01,Rice05} generated by spiral density waves.

The cooling rate of the disk (per side per unit surface area) is given by
\begin{equation}
F_{rad} = \frac{3}{8}\;\frac{\sigma T^4}{ (\tau + 2/3\tau)}\;,
\end{equation}
where $\tau = \kappa \Sigma/2$ is the optical depth of the disk.  This
expression allows one to switch smoothly from the optically thick $\tau\gg 1$
to the optically thin $\tau \ll 1$ radiative cooling limits. We approximate
the opacity coefficient $\kappa$ following Table 3 in the Appendix of
\cite{Bell94}. For the problem at hand, it is just the first four entries in
the Table are important as disk solutions with $T\simgt 2000 $ K are thermally
unstable\citep[see also Appendix B in][]{Thompson05} since opacity rises as
quickly as $\kappa \propto T^{10}$ in that region. This rather simple
approximation to the opacities is justified for the order of magnitude
parameter study that we intend to perform here. In addition, we set a minimum
temperature of $T=40$ K for our solutions. Even without any gas accretion,
realistic gas disks near galactic centres will be heated by external stellar
radiation to effective temperatures of this order or slightly larger. The main
conclusions of this paper do not sensitively depend on the exact value of the
minimum temperature or exact opacity law.

\begin{figure}
\centerline{\psfig{file=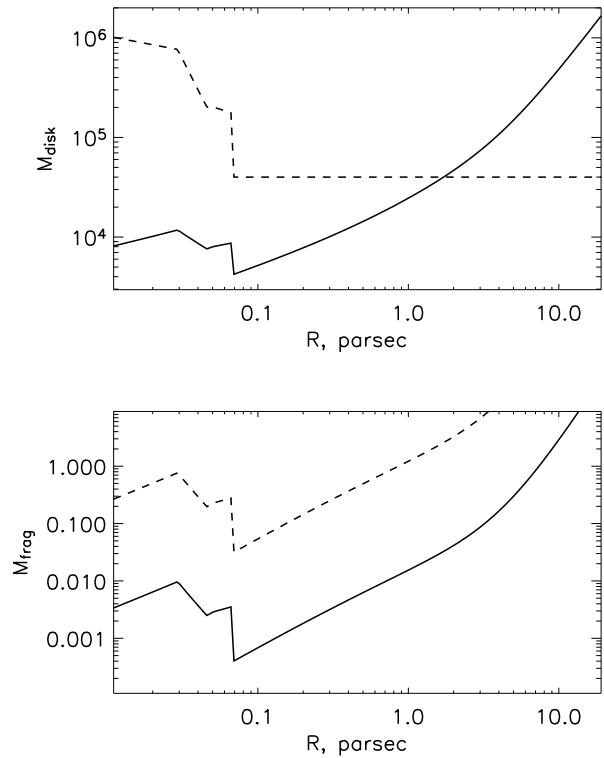,width=.49\textwidth}}
\caption{Disk mass (solid curve), $M_d = \pi \Sigma R^2$, and midplane
temperature (dashed) as a function of distance from the SMBH are shown in the
upper panel. The SMBH mass is that of \sgra. The lower panel shows two
estimates of the mass of the first fragments forming in the disk. The
realistic value of the fragments mass is likely to be in between these two
curves.}
\label{fig:fig1}
\end{figure}

\subsection{Masses of first stars in the disk}\label{sec:mfirst}

The upper panel of Figure \ref{fig:fig1} shows the resulting disk ``mass''
defined as $M_d = \pi \Sigma R^2$ and the midplane temperature (multiplied by
$10^3$).  The lower panel of the Figure shows two estimates of
mass of the first fragments in the disk. Different authors estimate the
volumes of the first unstable fragments slightly differently, but the
reasonable range seems to be from $H^3$ to $2H \times (2\pi H)^2$. The two
curves in the lower panel of Figure \ref{fig:fig1} should then encompass the
reasonable outcomes, from $M_{\rm frag} = \rho H^3$ to $M_{\rm frag} = \rho 8
\pi^2 H^3$. From the Figure, the fragment mass is, in the observationally
interesting range of radii, i.e. $R\sim 0.1-1$ pc, $M_{\rm frag}\simlt \msun$,
and hence if disk were to rapidly and completely collapse into clumps of mass
of this order, one would expect low-mass stars or even giant planets to
dominate the mass spectrum of collapsed objects.

 Numerical simulations with a constant cooling time show
 \citep[e.g.,][]{Gammie01} that if the disk cooling time is at the threshold
 for the fragmentation to take place, then the first gas clumps will grow very
 rapidly by inelastic collisions with other clumps, possibly until they reach
 the isolation mass $M_{\rm iso} \sim (\pi R^2 \Sigma)^{3/2}/\mbh^{1/2}$
 \citep{Levin03b}. If this is the case, then the main point of our paper --
 that stars born in an accretion disk near a SMBH are massive on average -- is
 proven, because the isolation mass can be hundreds to as much as $10^4$ Solar
 masses \citep{GoodmanTan04}. However we suspect that \cite{Gammie01}
 simulations yielded no further gravitational collapse of the gas clumps
 precisely because the cooling time were kept constant. As the clump density
 increases, the clump free-fall time decreases as $\propto \rho^{-1/2}$, and
 hence the clumps could not collapse as they could not cool rapidly enough. It
 is quite likely that had the cooling time {\em inside the clumps} were
 allowed to decrease as the clumps get hotter, the clumps would collapse
 before they agglomerate into larger ones.

\section{Effects of first stars on the disk}\label{sec:stars}

We shall now assume that gravitational instabilities in the $Q\approx 1$ disc
resulted in the formation of first proto-stars. According to the discussion in
\S \ref{sec:mfirst}, we conservatively assume that these proto-stars are low
mass objects, and show that in certain conditions even a small admixture of
these to the accretion disk may significantly affect its evolution.

\subsection{Coupling between stellar and gas disks}\label{sec:coupling}

As the stars are born out of the gas in a turbulent disc, we assume that the
initial stellar velocities are the sum of the bulk circular Keplerian velocity
$v_K$ in the azimuthal direction and a random component with three dimensional
dispersion magnitude $\sigma_0 \approx c_s$. This also implies that at least
initially stellar disk height-scale, $H_*$, is roughly the same as that of the
gas disk, $H$.  Proto-stars would interact by direct collisions and N-body
scatterings between themselves and also via dynamical friction with the gas.
The rate for proto-stellar collisions, $1/t_{\rm coll}$, is the sum of two
terms, the geometric cross-section of the colliding stars and the
gravitational focusing term \citep[e.g., see][]{Binney87}. One can show that
$1/t_{\rm coll} \Omega \simeq \hbox{max}[\Sigma_* R_{\rm coll}^2/M_*,
(\Sigma_*/\Sigma) R_{\rm coll}/H]$, from which it is obvious that collisions
are unimportant as long as the collision radius, $R_{\rm coll} \sim 2 R_{\rm
proto}$ (the proto-star radius), is much smaller than the disk height
scale. In all of the cases considered below this will be satisfied by few
orders of magnitude, therefore we shall neglect direct collisions.

The N-body evolution of the system of stars immersed into a gas disk is
described by \citep{NC05}
\begin{equation} 
\frac{d \sigma}{d t} \sim 4 \pi G^2 M_* \left[ \frac{\rho_* \ln
\Lambda_*}{\sigma^2} - \frac{\rho C_{\rm d} \sigma}{(c_{\rm s}^2 +
\sigma^2)^{3/2}}\right]\;
\label{dsnet}
\end{equation}
where $\ln \Lambda*\sim$~few is the Coulomb logarithm for stellar collisions,
$C_{\rm d}\sim$~few is the drag coefficient for star-gas interactions
\citep{Artymowicz94}, $\sigma$ is one-dimensional velocity dispersion, and
$\rho_* = \Sigma_*/2 H_*$ is the stellar surface density. Therefore, as long
as the gas density $\rho \simgt \rho_*$, stellar velocity dispersion cannot
grow as it is damped by interactions with the gas too efficiently. Recalling
that $\rho_* = \Sigma_*/2H_*$, we find that in this situation
\begin{equation} 
\frac{\sigma}{c_s} \approx \frac{H_*}{H} \sim
\left(\frac{\Sigma_*}{\Sigma}\right)^{1/4} < 1\;.
\end{equation}
Thus, initially, when $\Sigma_*\ll \Sigma$, stars are embedded in the gaseous
disk and form a disk geometrically thinner than that of the gas. However, if
stellar surface density grows and approaches that of the gaseous component,
then the stellar velocity dispersion will run away. The stars then form a
geometrically thicker disk (numerical simulations, to be reported in a future
paper, confirm these predictions). Galaxy disks apparently operate in this
regime, with molecular gas having a much smaller scale height than stars.

\subsection{Accretion onto proto-stars}

Since the stars remain embedded in the disk, they will continue to gain mass
via gas accretion. Such accretion has been previously considered by many
authors \citep[e.g.,][]{Lissauer87,Bate03,GoodmanTan04}. We assume that the
accretion rate is 
\begin{equation}
\dot M_* = \hbox{min}\left[\dot M_{\rm Bondi}, \dot M_{\rm Hill}, \dot M_{\rm
Edd}\right]\;,
\label{mdot}
\end{equation}
where the accretion rates in the brackets are the Bondi, the Hill and the
Eddington limit, respectively \citep[e.g.][]{Nayakshin05}. The latter is
calculated based on the Thomson opacity of free electrons instead of dust
opacity because we assume that the cooler regions of accretion flow onto the
star are shielded from the stellar radiation by the inner, hotter accretion
flow \citep{Krumholz05}: $\dot M_{\rm Edd} = 10^{-3} r_* \msun$ year$^{-1}$,
where $r_* = R_*/\rsun$.

\subsection{Heating of the disc by proto-stars}

Presence of the stars will lead to additional disc heating via radiation and
outflows, and N-body scattering. The energy liberation rate per surface area
due to N-body interactions is given by
\begin{equation}
Q^+_{*N} \sim  \Sigma_* \sigma \left(\frac{d \sigma}{d t}\right)_* \sim 4 \pi G^2
M_* \Sigma_* \frac{\rho_* \ln \Lambda_*}{\sigma}\;,
\end{equation}
where $(d\sigma/dt)_*$ stands for the first term only in equation
\ref{dsnet}. Internal disk heating with the $\alpha$-parameter equal to unity is 
\begin{equation}
Q^+_{\rm d} = \frac{9}{4} \Sigma c_s^2 \Omega\;.
\label{hvisc}
\end{equation}
Comparing the stellar N-body heating with that of the internal disk heating,
we have
\begin{equation}
\frac{Q^+_{*N}}{Q^+_{\rm d}} \sim \left(\frac{\Sigma_*}{\Sigma}\right)^{3/2}
\frac{M_*}{\mbh}\; \left(\frac{R}{H}\right)^3
\end{equation}
We have assumed above that Toomre-parameter $Q\sim 1$ when the stars just
appear in the disk, and that $\Omega^2 = G\mbh/R^3$, i.e. that we are within
the SMBH sphere of influence.  Considering this expression for typical
numbers, one notices that N-body heating is never important for large black
holes and disks with finite disk thickness, i.e. at distances of tens of
parsec and further away, but it may become important for smaller SMBH such as
\sgra\ and sub-parsec distances.

Radiative internal output of the stars is another source of disk heating. We
shall use a very simple parameterisation for the internal stellar luminosity
as a function of mass, $L_* \propto M_*^3$. To this radiative output we should
also add the accretion luminosity:
\begin{equation}
L_{\rm acc} = \frac{G M \dot M_*}{R_*}\;.
\label{lacc}
\end{equation}
 The sum should not exceed the Eddington limit, $\ledd = 4\pi G M_* m_p c/
 \sigma_T \approx 10^{38} m_*$~erg s$^{-1}$, and hence our prescription is
\begin{equation}
L_* = \hbox{min}\left[\lsun \frac{M_*^3}{\msun^3} + L_{\rm acc},
\ledd\right]\;,
\label{lstar}
\end{equation}
where $\ledd = 4\pi G M_* m_p c/ \sigma_T$ is the Eddington limit with Thomson
opacity $\sigma_T$. 
The radiative disk heating per unit surface area is then
\begin{equation}
Q^+_{*\rm rad} = \frac{\Sigma_*}{M_*} L_*\;.
\label{qstar}
\end{equation}

\section{Analytical estimates}\label{sec:analytical}

As equation \ref{rho} suggests, accretion disks on parsec scales are as dense
as $10^{12}$ particles cm$^{-3}$, which is multiple orders of magnitude denser
than the densest gas in molecular clouds far from galactic centers. Therefore
it is not surprising that the first proto-stars will be accreting at
super-Eddington rates, typically. The corresponding accretion luminosity
heating (equation \ref{qstar} with $L_* = \ledd$) is
\begin{equation}
Q^+_{*\rm rad} \sim 10^5  \; \Sigma_* r_* \; \hbox{erg s}^{-1} \hbox{cm}^{-2}\;.
\label{qapprox}
\end{equation}
At the same time, the disk intrinsic heating at $Q\approx 1$ is, from equation
\ref{hvisc},
\begin{equation}
Q^+_{\rm d} \approx \; \Sigma \; T_2 \; \frac{\mbh}{3\times 10^6 \msun } \;
\left[\frac{0.1 \hbox{pc}}{R}\right]^{3/2} \; \hbox{erg s}^{-1}
\hbox{cm}^{-2} \;,
\label{qself}
\end{equation}
where $\Sigma$ and $\Sigma_*$ are in units of g cm$^{-2}$, and $T_2$ is the
disk temperature in units of 100 K. We see that even a very small admixture of
proto-stars ($\Sigma_* \ll \Sigma$) accreting at Eddington accretion rates will
result in stellar heating much exceeding the intrinsic one. Since the disk
thermal equilibrium is established on time scales comparable to 
the disk dynamical time, the disk will heat up at a constant $\Sigma$ until its
radiative losses can balance the accretion luminosity. This will increase the
disk sound speed and the \cite{Toomre64} Q-parameter {\em above
unity}. Therefore, the accretion feedback will stop further fragmentation from
happening. The stars embedded into the disk will however continue to gain mass
at very high rates. This should lead to a top-heavy initial mass function for
the stars. 

Note that a similar conclusion has been already reached by \cite{Levin03b} who
considered rather later stages in the evolution of a more massive AGN disk,
when the high mass stars were turned into stellar mass black holes. He pointed
out that accretion onto these embedded black holes will likely heat the disk,
driving the \cite{Toomre64} $Q$-parameter above unity for radii somewhat
smaller than a parsec.

\section{Numerical results}\label{sec:post}

\begin{figure}
\centerline{\psfig{file=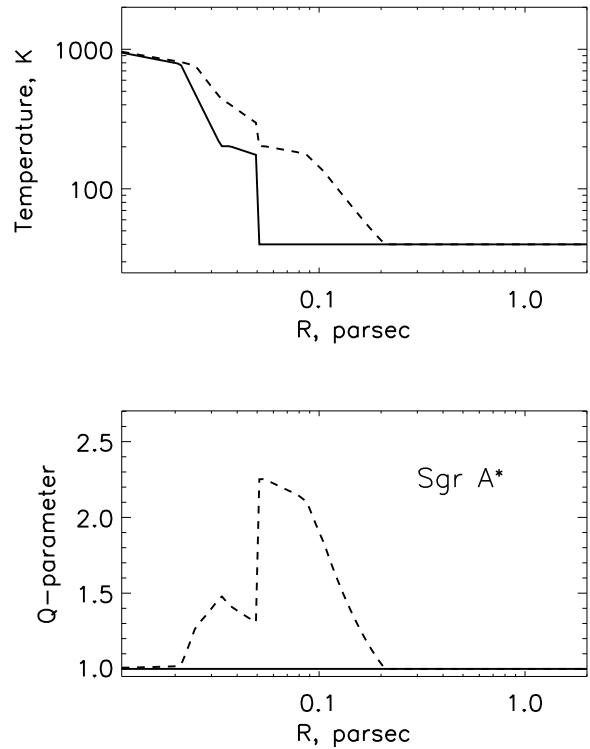,width=.49\textwidth}}
\caption{Temperature (upper panel) and $Q$-parameter (lower panel) versus
  radius in parsec for the marginally self-gravitating disk with (dashed) and
  without (solid) proto-stars embedded in the disks. The proto-stars have
  masses $M_*= 0.1 \msun$ and surface density $\Sigma_* = 0.001 \Sigma$.}
\label{fig:fig2}
\end{figure}

We shall first consider the case of \sgra\ for which the mass of the SMBH is
estimated to be $\mbh \simeq 3.5\times 10^6\msun$
\citep{Schoedel02,Ghez03a}. For the particular example, we shall accept that
$\Sigma_* = 0.001 \Sigma$ and that the initial masses of proto-stars are $M_*
= 0.1\msun$. The upper panel of Figure \ref{fig:fig2} shows the disk midplane
temperature before the stars are introduced (solid) and after (dashed) versus
radius $R$. Temperature of the gas increases for radii $0.03 \simlt R \simlt
0.2$ parsec. \cite{Toomre64} Q-parameter after the stars are introduced is
plotted in the bottom panel of Figure \ref{fig:fig1}. $Q$ indeed becomes
greater than unity in the same radial range, thus shutting off further
gravitational collapse.

\begin{figure}
\centerline{\psfig{file=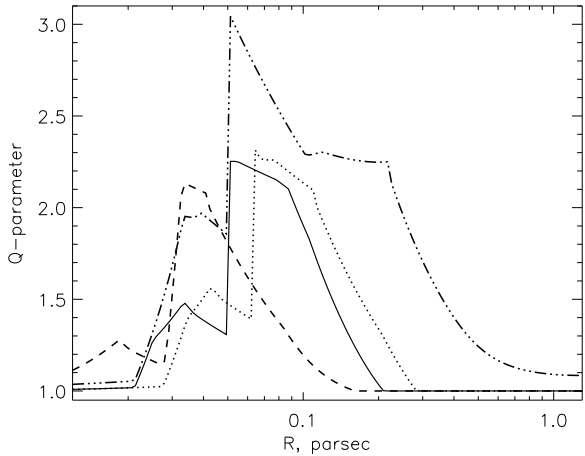,width=.49\textwidth}}
\caption{Toomre $Q$-parameter versus radius for the self-gravitating disk in
  which first stars were born. The solid line is the same as that in Figure
  \ref{fig:fig2}, lower panel. The other curves are obtained by varying
  assumption of the model to test sensitivity of the results (see text).}
\label{fig:fig3}
\end{figure}

The radial range where the proto-stars shut off further fragmentation is
rather insensitive to assumptions of our model. Figure \ref{fig:fig3} shows
the $Q$-parameter after stars are introduced into the $Q=1$ gaseous disk for
\sgra\ case but with varying assumptions. In particular, the solid curve is
the same as that in Figure \ref{fig:fig2}, lower panel; the dotted one is
calculated for the opacity coefficient $\kappa$ multiplied arbitrarily by 3,
whereas for the dashed one $\kappa$ was divided by $\sqrt{T}$. These arbitrary
changes were introduced to estimate the degree to which the results are
dependent on the (uncertain) opacity detail. Finally, the dot-dashed curve is
calculated assumed the standard opacity but increasing the proto-stellar mass
to $1\msun$ and stellar surface density $\Sigma_*$ to 0.01, respectively. The
stellar heating is then more pronounced and a larger area of the marginally
stable disk can be affected.

\begin{figure}
\centerline{\psfig{file=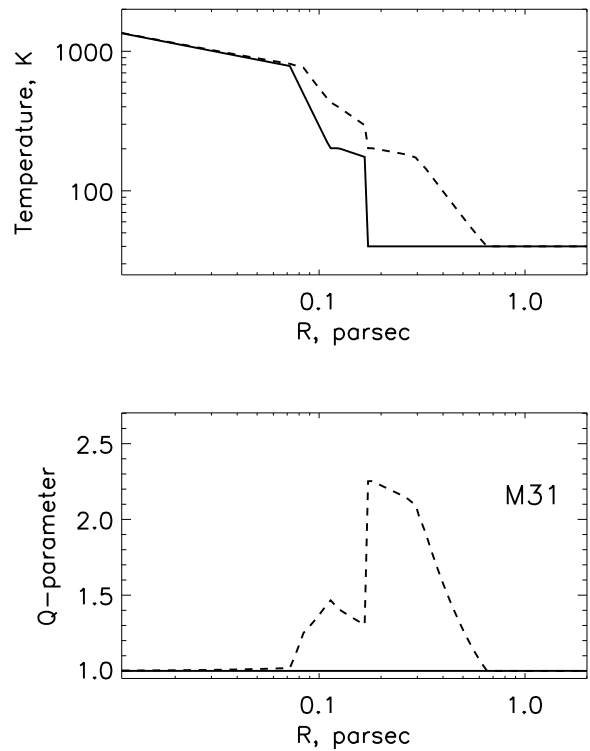,width=.49\textwidth}}
\caption{Same model as that used for Figure \ref{fig:fig1}, except for a
  higher SMBH mass, $\mbh = 1.4\times 10^8 \msun$. Note that, compared with
  Figure \ref{fig:fig1}, the radial range where proto-stars succeed in heating
  the disk up has shifted to slightly larger radii.}
\label{fig:m31}
\end{figure}

We also consider the case of a more massive black hole, in particular we set
$\mbh = 1.4\times 10^8 \msun$, as thought to be the case for M31. Figure
\ref{fig:m31} shows the disk temperature structure (upper panel) before the
collapse (dashed) and after the collapse. Note that these curves are almost
identical to those for \sgra\ case except for a general shift to larger
radii. This shift is about a factor of 3 only, which should not be surprising
given that in the standard accretion disk theory \citep{Shakura73} the
midplane disk temperature is a very weak function of the central object mass.

\section{Discussion}

In this semi-analytical paper, we studied the ``first minutes'' of an
accretion disk around a super-massive black hole after the disk became
unstable and formed first stars. We assumed that the disk accumulated its mass
over time scales much longer than the local dynamical time, and is thus in a
thermal equilibrium before the gravitational collapse. In this case,
irrespectively of the typical mass of the first proto-stars, even a 0.1\%
admixture (by mass) of these significantly alters the thermal energy balance
of the disk. The proto-stars accrete gas from the surrounding disk at
very high (super-Eddington) rates at a range of disk radii. The accretion
luminosity of these stars is sufficient to heat the disk up in that range of
radii to the point where it becomes stable to self-gravity ($Q > 1$), which
then shuts off further fragmentation of the disk. The proto-stars already
present in the disc would however continue to gain mass at very high
rates. Quite generally, then, an average star created in such a disk will be a
massive one, in stark contrast to the typical galactic star formation event.

Significance of accretion feedback onto embedded {\em stellar mass black
holes} for accretion disks near galactic centres was pointed out by
\cite{Levin03b}. He noted that the accretion disks can be stabilised by the
feedback out to radius of about one parsec, in good agreement with our
results. Since the Eddington luminosity depends only on the disk opacity and
the mass of the central object, it is then not really surprising that the
feedback is effective for accretion onto stars as well.

The range of radial distances from the SMBH where this effect operates is a
slow function of the SMBH mass, and is typically from a fraction of 0.1 parsec
to a few parsec, with the peak of the effect taking place at $R\sim 0.1$
parsec. The reason why the feedback is only effective in a range of radii is
that at large radii, i.e. tens of parsec, the gas density in the disk drops
significantly so that accretion rates onto the proto-stars become much smaller
than the respective Eddington-limited rates. At radii much smaller than $\sim
0.1$ parsec, the intrinsic disk heating (equation \ref{qself}) becomes very
large. A related point is that steady-state constant accretion rate disk
models show that there is always the minimum radius where star formation
becomes impossible as $Q > 1$ there
\citep[e.g.][]{Goodman03,Levin03b,NC05}. The value of the minimum distance
where star formation should be expected is comparable to the minimum radius
for which we predict favourable conditions for development of a top-heavy IMF.

We suggest that the uncommonly effective feedback from star formation on
low-mass proto-stars may be relevant for the formation of rings of massive
stars observed in the Galactic Centre and in the nucleus of M31. In
particular, \cite{Nayakshin05b,NS05} have shown from two completely
independent lines of evidence that the IMF of stellar disks in the Galactic
Centre must have been very top-heavy, with the mass of Solar-type stars
accounting for no more than $\sim 50$\% of the total, the remainder being in
the $M\simgt 30 \msun$ stars.  The insensitivity of our results to details of
the model and the SMBH mass suggest that the IMF of stars born inside
accretion disks near galactic centers may be generically top-heavy, which
would have long-ranging consequences for the accretion theory in AGN.

\subsection{Generality and shortcomings of this work}

In this paper we concentrated on the growth of proto-stellar mass via
accretion of gas. As discussed in \S \ref{sec:stars}, in certain conditions
the mass of gas clumps before they collapse to form a star may be much higher
than the Jeans mass. This would only increase the expected final mass of a
typical star in the disk. The same is true for direct collisions of
proto-stars, the other channel via which proto-stars may grow (see \S
\ref{sec:coupling}).

In addition, we have here limited the rate at which the proto-star would grow
to the Eddington accretion rate onto a star. This is important if the rate at
which the gas is captured in the sphere of influence of the proto-star, the
Hill or the Bondi radius, whichever is smaller, exceeds the Eddington accretion
rate. It is possible that in reality the excess gas settles into a
rotationally supported ``proto-stellar'' disk from which further generations
of stars may be born \citep{Milosavljevic04}. It is not obviously clear
whether this effect would increase the average mass of the stars or would
rather decrease it. On the one hand, fragmentation of the proto-stellar disk
may give rise to many low mass stars. On the other, though, these stars may be
then driven into the central more massive star by the continuing gas
accretion, as suggested by \cite{Bonnel05}. In the latter case the central
star may in fact grow faster than the Eddington accretion rate.

On the balance, we believe that the main conclusion of our work, e.g. the
unusually high (perhaps dominant) fraction of the total mass going into
creation of high mass stars as opposed to low mass stars, may be rather
robust. One clear exception to this will be a very rapid (dynamical)
gravitational collapse of a disc. For example, when a large quantity of gas
(compared to the minimum needed for the disk to become self-gravitating, see
Figure 1) cools off very rapidly and settles into a disc configuration, and
the cooling time is shorter than dynamical time, the disc will break into
self-gravitating low mass objects before it can establish thermal balance
\citep[e.g.,][]{Shlosman89}.

We deliberately stayed away here from discussing the much more complicated
question of the eventual disk evolution. The answer depends not only on the
initial radial structure of the disk but also on how the disk is fed with gas
after it crossed the self-gravity instability threshold. We shall investigate
these issues in our future work (note that Thompson et al. 2005 recently
developed a model for kilo-parsec scale star-forming disks in ultra-luminous
galaxies).

\subsection{Why different from ``normal'' star formation?}\label{sec:normal}

It is instructive to emphasise the differences in star formation rates near a
SMBH and in a galaxy.  Consider the relevant gravitational collapse time
scales $t_{\rm c}$. For accretion disks around the SMBH, this is $t_{\rm c}
\simeq 1/\Omega \approx 60\; \hbox{years} \; (R/0.04 \;\hbox{parsec})^{3/2} \;
(\mbh/3\times 10^6 \msun)^{-1}$, which is shorter than the Eddington limit
doubling time of $\sim$ a thousand years. Compare this time with the free-fall
time for a molecular cloud of mass $10^3\msun$ and size of 1 parsec: $t_{\rm
ff} \sim 10^6$ years. Clearly, then, an average accretion rate in the galactic
environment is orders of magnitude below the Eddington limit, and no
significant radiation feedback should be expected from {\em low mass}
proto-stars. The latter can then form in great numbers with little damage to
the rest of the cloud, unlike in the case of a massive disk.

Another significant qualitative difference is that the escape temperatures,
$T_{\rm esc} \sim GM\mu/kR$, are vastly different near a SMBH and inside a
molecular cloud. For the former, it is typically in the range of $10^6 - 10^7$
K, whereas for the latter it is only $\sim$ few $\times 10^3$ K. Hence, while
photo-ionizing feedback from massive stars may unbind most of the gas in a
molecular cloud, stopping not only further fragmentation but also further
accretional growth of proto-stars, in SMBH disks the effect is local
\citep[e.g.,][]{Milosavljevic04}. In particular, it simply increases the disk
scale-height until the gravity of the SMBH \citep[which increases as $z/H$ for
thin accretion disks, e.g.,][]{Shakura73} is strong enough to hold the gas in
place. In other words, accretion or any other star formation feedback in a
disk environment may be strong enough to prevent further disk fragmentation but
not the growth of the existing proto-stars.

Third important difference is geometry. Most stars in a molecular cloud move
on orbits different from those of the gas, as the latter is influenced by both
gravity and pressure forces whereas stars obey only the gravity.  Hence the
gas and the stars may be separated out in space, terminating accretion onto
the stars. In contrast, as is well known from the standard accretion theory
\citep{Shakura73}, gas pressure forces are very small compared to the SMBH
gravity for thin gas disks in galactic centers, and so both stars and gas
follow essentially circular Keplerian orbits around the SMBH
\citep{NC05}. Therefore the stars are always not too far away from the gas and
hence have a much better chance to gain more mass by accretion.

\section{Conclusions}

In this paper we have shown that the birth of even a small number (by mass
fraction) of low-mass proto-stars inside a marginally stable accretion disk
near a galactic center will unleash a very strong thermal feedback onto the
gaseous disk. In a sub-parsec range of radii, the disk will be heated and
thickened so that it becomes stable to further fragmentation. The feedback
however is not strong enough to unbind the gas from the deep potential well of
the SMBH. Therefore, while the feedback stops a further disk fragmentation,
accretional growth of stars already present in the disk proceed. Quite
generically, this scenario should lead to the average star created in the SMBH
accretion disk being ``obese'' compared to its galactic cousins.

The author acknowledges very useful comments on the draft by Yuri Levin, and
fruitful discussions with Andrew King, Jim Pringle and Jorge Cuadra.

\end{document}